\documentclass[aps,pra,showkeys,twocolumn,superscriptaddress]{revtex4-2}
\usepackage{array}
\usepackage{booktabs}
\usepackage{tabu}
\usepackage{dcolumn}
\usepackage{floatrow}
\usepackage{amsmath}
\usepackage{amsfonts}
\usepackage{float}
\usepackage{amssymb}
\usepackage{graphicx,color}
\usepackage[colorlinks={true}]{hyperref}
\hypersetup{colorlinks=true,linkcolor=blue,citecolor=blue,urlcolor=blue}
\usepackage{graphicx}
\usepackage{dcolumn}
\usepackage{bm}
\usepackage{pstricks}
\usepackage{braket}
\usepackage{orcidlink}
\usepackage{lipsum}
\def\be{\begin{equation}}
  \def\ee{\end{equation}}
\def\bea{\begin{eqnarray}}
\def\eea{\end{eqnarray}}
\def\f{\frac}
\def\n{\nonumber}
\def\l{\label}
\def\p{\phi}
\def\o{\over}
\def\R{\rho}
\def\pa{\partial}
\def\om{\omega}
\def\na{\nabla}
\def\P{\Phi}
\begin{document}

\title{Spatial Phase Control of Energy and Ergotropy in Quantum Batteries} 

\author{Maryam Hadipour \orcidlink{0000-0002-6573-9960}}
\affiliation{Faculty of Physics, Urmia University of Technology, Urmia, Iran}

\author{Soroush Haseli \orcidlink{0000-0003-1031-4815}}\email{soroush.haseli@gmail.com}
\affiliation{Faculty of Physics, Urmia University of Technology, Urmia, Iran}

\date{\today}
\def\be{\begin{equation}}
  \def\ee{\end{equation}}
\def\bea{\begin{eqnarray}}
\def\eea{\end{eqnarray}}
\def\f{\frac}
\def\n{\nonumber}
\def\l{\label}
\def\p{\phi}
\def\o{\over}
\def\R{\rho}
\def\pa{\partial}
\def\om{\omega}
\def\na{\nabla}
\def\P{$\Phi$}

\begin{abstract}
We investigate the role of spatial geometry in controlling energy storage and work extraction in a non-Markovian quantum battery. The system consists of two identical two-level systems coupled to a structured waveguide reservoir, where one qubit serves as the charger and the other as the battery. The spatial separation between the qubits introduces a geometry-dependent phase that governs collective interference, directly modulating the decay channels and energy transfer dynamics. We show that tuning the interqubit distance allows for the engineering of bright and dark collective pathways, which, in conjunction with environmental memory, significantly enhances the charging power, maximal stored energy, and ergotropy revivals.
\end{abstract}


\maketitle

\section{Introduction}

The emerging field of quantum thermodynamics provides a rigorous framework for understanding energy conversion, work extraction, and dissipation at the nanoscale~\cite{a1,a2,a3,a4,a5,a6,a7}. Within this paradigm, quantum batteries (QBs) have attracted substantial attention as non-classical energy storage platforms capable of outperforming their classical counterparts~\cite{a8,r1,r2,r3,r4,r5,r6,a9,a10}. By leveraging non-classical resources such as quantum coherence and multipartite entanglement, quantum batteries can achieve cooperative charging speedups and enhanced power output~\cite{a11,a12,a13,a14,a15,a16}. Diverse physical implementations have been proposed to realize these thermodynamic advantages, ranging from coupled spin chains~\cite{a17,a18} and cavity or circuit quantum electrodynamics setups~\cite{a19,a20,a21} to trapped ions and quantum dot configurations~\cite{a22,a23,a24}.

In realistic physical scenarios, however, any quantum battery unavoidably interacts with its surrounding environment \cite{b0}. This coupling induces decoherence, energy dissipation, and unwanted self-discharge, posing a major bottleneck to stable energy retention and coherent work extraction~\cite{a25,a26,a27,a28,a29,a30}. To mitigate these detrimental environmental effects, considerable effort has focused on exploiting structured reservoirs and non-Markovian memory. Memory effects can promote information and energy backflow from the reservoir into the system, enabling partial revivals of stored energy~\cite{a31,a32,a33,b1,b2,a34,a35}. Concurrently, collective dissipation mechanisms, such as subradiant or dark states, offer passive protection against environmental decay~\cite{b3,a36,a37,a38}. Despite these advances, identifying flexible physical mechanisms to selectively turn dissipative channels on or off and protect genuinely extractable work remains an open challenge. Waveguide quantum electrodynamics (QED) provides a compelling platform to address this issue. In one-dimensional structured waveguides, mediated interactions exhibit long-range coherence, while mirror boundaries establish standing-wave field profiles with strong spatial dependence. Consequently, the position of the quantum emitters relative to the standing-wave nodes and antinodes provides a convenient, continuously tunable spatial phase that acts as an effective control parameter. While spatial separation is recognized for tailoring superradiant and subradiant collective emission in multi-emitter arrays~\cite{a39,a40}, its direct potential for optimizing quantum thermodynamic figures of merit, particularly the distinction between raw stored energy and coherently extractable ergotropy, has not been systematically examined in non-Markovian waveguide environments. In this work, we investigate the spatial phase control of energy, ergotropy, and charging power in a two-qubit quantum battery system coupled to a mirror-terminated structured waveguide. By treating one qubit as an initially excited charger and the second as the battery, we demonstrate that tuning their relative positions and boundary distances establishes destructive or constructive interference across the symmetric and antisymmetric collective dissipation channels. We analyze how this geometric interference cooperates with non-Markovian reservoir memory to suppress radiative leakage, sustain stable population inversion, and optimize transient work extraction. In particular, we show that charging power is markedly sensitive to reservoir bandwidth, whereas direct coherent coupling enhances the sensitivity of peak thermodynamic metrics to emitter positioning. 
The work is organized as follows. In Sec.~\ref{sec:measures}, we review the thermodynamic figures of merit used to evaluate charging performance, focusing on stored energy, ergotropy, and charging power. Section~\ref{sec:model} presents the physical model of the charger-battery system coupled to the structured waveguide, detailing the spatial phase-dependent collective dissipation channels and the non-Markovian reservoir characteristics. Finally, Sec.~\ref{sec:conclusion} summarizes the main conclusion.

\section{Figures  of merit}\label{sec:measures}
In this section, we present the thermodynamic framework employed to evaluate the performance of the quantum battery.
 We consider a battery modeled by the Hamiltonian $H_B$ and characterized by the density operator $\rho_B$. The spectral decomposition of the battery state takes the form
\begin{equation}
\rho_B = \sum_j r_j |r_j \rangle \langle r_j |,
\end{equation}
where the eigenvalues $r_j$ are arranged in descending order, $r_j \ge r_{j+1}$, with $|r_j \rangle$ representing the corresponding orthonormal eigenvectors, which naturally satisfy the normalization condition $\sum_j r_j = 1$. In the same manner, the battery Hamiltonian can be expanded in its energy eigenbasis as
\begin{equation}
H_B = \sum_i \mathcal{E}_i |\mathcal{E}_i \rangle \langle \mathcal{E}_i |,
\end{equation}
where the energy eigenvalues are labeled in non-decreasing order, $\mathcal{E}_i \leq \mathcal{E}_{i+1}$, with $|\mathcal{E}_i\rangle$ denoting the associated eigenstates. The energy stored in the battery is given by the expectation value of its Hamiltonian,
\begin{equation}
E_B = \operatorname{tr}(\rho_B H_B).
\end{equation} 
To quantify the fraction of stored energy that can be converted into useful work, we employ ergotropy, which measures the maximum work extractable via cyclic unitary operations~\cite{a35}. This bound is determined by comparing the energy of the actual state with that of its associated passive counterpart $\sigma_\rho$, a state from which no further work can be unitarily extracted, yielding
\begin{equation}\label{erg1}
\mathcal{W} = \operatorname{tr}(\rho_B H_B) - \operatorname{tr}(\sigma_\rho H_B).
\end{equation}
The passive state $\sigma_\rho$ is diagonal in the energy eigenbasis and contains no population inversion, reading
\begin{equation}
\sigma_\rho = \sum_i r_i |\mathcal{E}_i \rangle \langle \mathcal{E}_i |,
\end{equation}
where the populations $r_i$ are ordered monotonically decreasing and paired with increasing energy eigenvalues $\mathcal{E}_i$. A defining advantage of quantum batteries lies in their capacity for rapid charging, a feature typically evaluated in terms of either instantaneous or average charging power. As a practical measure of performance, the average charging power is defined as the ratio of the energy stored in the battery to the charging duration, given by

\begin{equation}
\mathcal{P}_{\text{avg}}(t) = \frac{E_B(t)}{t}.
\end{equation}
Complementing this measure, the instantaneous charging power quantifies the rate of energy accumulation at any given time, defined as
$
\mathcal{P}(t) = \frac{d E_B(t)}{dt}
$. This work aims to uncover how spatial geometry governs the charging and work-extraction dynamics of a quantum battery. Accordingly, the relative arrangement of the charger and battery serves here as an active control parameter. The following section lays out the theoretical model and details the underlying physical framework.

\section{Physical model}\label{sec:model}
\begin{figure}
\centering
\includegraphics[width=\textwidth]{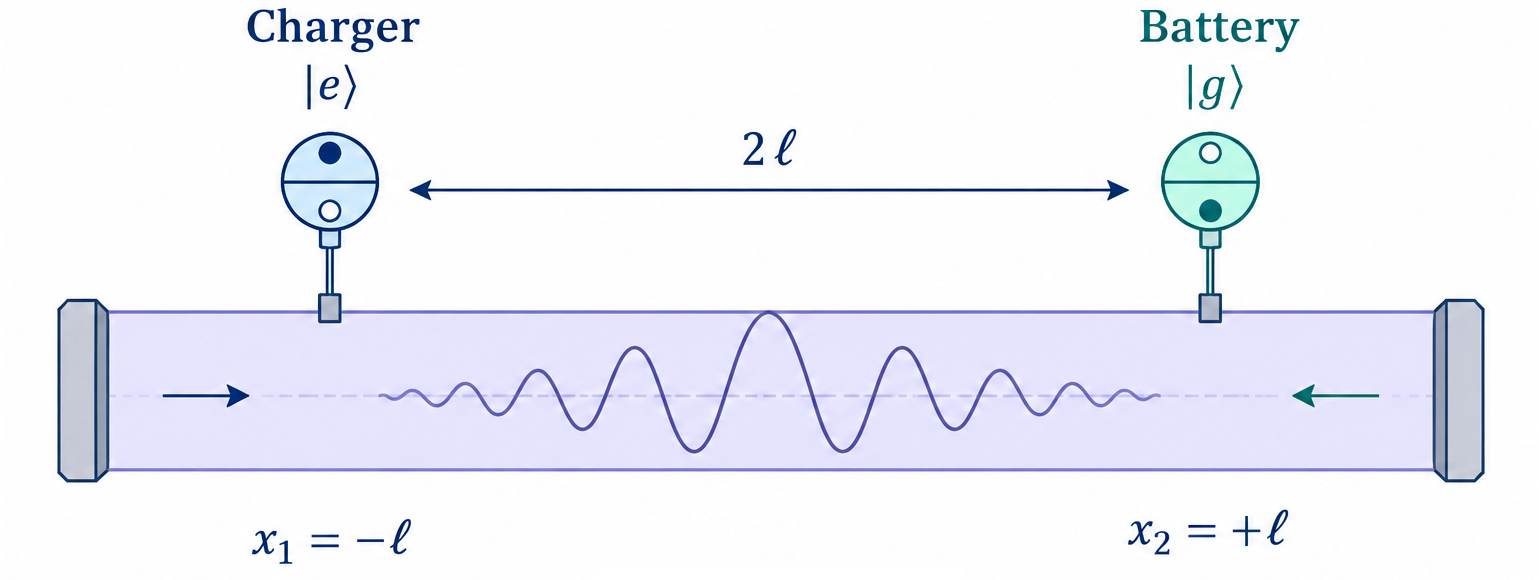}
\caption{Schematic of the physical model. Two identical qubits are separated by a distance $2l$ inside a mirror-terminated one-dimensional waveguide. Qubit 1 acts as the charger (initially in the excited state), while qubit 2 acts as the quantum battery (initially in the ground state).}
\label{Fig1}
\end{figure}

We consider a composite quantum system of two identical qubits located at positions $x_1 = -l$ and $x_2 = +l$ within a mirror-terminated one-dimensional waveguide of length $L$. In this quantum battery framework, one qubit operates as the charger while the other acts as the battery
\cite{b3}. A schematic illustration of this configuration is depicted in Fig.~\ref{Fig1}. The total Hamiltonian of the composite system is given by
$
H = H_S + H_E + H_I
$,
where $H_S$ denotes the bare Hamiltonian of the charger and battery qubits, $H_E$ describes the bosonic modes of the structured environment, and $H_I$ accounts for the light-matter interaction coupling the two subsystems. The free Hamiltonian of the system is given by~\cite{b3}
\begin{equation}
H_S = \omega_0 \sum_{i=1}^{2} \sigma_i^+ \sigma_i^- + \zeta (\sigma_1^+ \sigma_2^- + \sigma_1^- \sigma_2^+),
\end{equation}
where $\omega_0$ represents the transition frequency of the two-level systems, and $\zeta$ denotes the direct dipole-dipole coupling strength between the qubits~\cite{a36}.

Within our framework, qubit 1 serves as the charger and qubit 2 as the battery, where $\sigma_j^{\pm}$ denote the standard raising and lowering operators for the $j$-th qubit. The Hamiltonian of the environmenta  is given by
$
H_E = \sum_k \omega_k a_k^\dagger a_k,
$,
where $a_k^\dagger$ and $a_k$ denote the creation and annihilation operators for the $k$-th standing-wave mode of the waveguide, respectively. Under the rotating-wave approximation, the interaction Hamiltonian is given by
\begin{equation}\label{hamilton}
H_I = \sum_k \left[ \xi_k^{(1)} \sigma_1^- a_k^\dagger + \xi_k^{(2)} \sigma_2^- a_k^\dagger + \text{h.c.} \right],
\end{equation}
where $\xi_k^{(j)}$ quantifies the coupling strength between the $j$-th qubit and the $k$-th environmental mode, and $\text{h.c.}$ denotes the Hermitian conjugate. Geometric effects are manifested through the position-dependent coupling amplitudes $\xi_k^{(j)} = \xi_k e^{i k x_j}$, where $x_j$ denotes the coordinate of the $j$-th qubit and $k$ represents the wave number~\cite{a37,a38,a39,a40}. Throughout this work, we set $\hbar = 1$ and adopt both the dipole and rotating-wave approximations

Because the system dynamics are primarily governed by environmental modes near the qubit transition frequency $\omega_0$, we evaluate the wave number in the position-dependent phase terms at the resonant value $k \approx k_0 = 2\pi/\lambda_0$, where $\lambda_0$ denotes the corresponding resonant wavelength. Under this approximation, the position-dependent coupling coefficients simplify to~\cite{a41,a42}
\begin{equation}
\xi_k^{(1)} \simeq \xi_k e^{-i \theta}, \quad \xi_k^{(2)} \simeq \xi_k e^{+i \theta}.
\end{equation}
where $\theta=k_0 l$ is spatial phase.  To simplify the dynamics, it is convenient to introduce the collective symmetric $|S\rangle = \frac{1}{\sqrt{2}} \left( |e_C g_B \rangle + |g_C e_B \rangle \right)$ and antisymmetric $|A\rangle = \frac{1}{\sqrt{2}} \left( |e_C g_B \rangle - |g_C e_B \rangle \right)$ basis states, where $|e\rangle$ and $|g\rangle$ denote the excited and ground states of the respective subsystems, with subscripts $C$ and $B$ referring to the charger and battery.

In this basis, the dynamics decouple into independent channels corresponding to symmetric and antisymmetric contributions, thereby simplifying the mathematical treatment. Furthermore, this transformation provides a clear physical insight into the energy transfer process, explicitly highlighting the role of coherent and dissipative interactions in the quantum battery's charging dynamics. In this representation, and within the near-resonant approximation, the interaction Hamiltonian $H_I$ can be compactly written as
\begin{equation}\label{HI}
H_I = \sum_k \sqrt{2} 
\left(
\Gamma_{s,k} \, \Xi_s^{+} 
+ \Gamma_{a,k} \, \Xi_a^{+}
\right) a_k 
+ \mathrm{H.c.},
\end{equation}
where the geometry-dependent coupling strengths are given by
\begin{equation}
\Gamma_{s,k} = \xi_k \cos\theta, \qquad \Gamma_{a,k} = -i \xi_k \sin \theta,
\end{equation}
and the collective ladder operators are defined as $\Xi_s^{+} = \frac{\sigma_1^{+} + \sigma_2^{+}}{\sqrt{2}}$ and $\Xi_a^{+} = \frac{\sigma_1^{+} - \sigma_2^{+}}{\sqrt{2}}$.
This representation explicitly illustrates how the spatial geometry dictates the system dynamics via the phase factor $\theta$, tuning the effective coupling of the symmetric and antisymmetric collective channels to the environment. As evident from Eq.~\eqref{HI}, the symmetric and antisymmetric collective modes couple to the reservoir via independent channels whose strengths are governed by the interqubit distance $2l$. Because the total excitation number is conserved under the rotating-wave approximation, and assuming the reservoir is initially in the vacuum state $|0\rangle_E$, the dynamics is strictly confined to the single-excitation subspace. Accordingly, the initial state of the total system-environment composite is expressed as
\begin{equation}\label{eq:initial_state}
|\psi(0)\rangle =
\left[
c_{1}(0)|e_{c}, g_{B}\rangle
+
c_{2}(0)|g_{c}, e_{B}\rangle
\right]
\otimes |0\rangle_{E},
\end{equation}
where $|g_{c}\rangle$ ($|g_{B}\rangle$) and $|e_{c}\rangle$ ($|e_{B}\rangle$) denote the ground and excited states of the charger (battery), respectively, and $|0\rangle_{E} = \bigotimes_k |0_k\rangle$ represents the multimode vacuum state of the environment. The initial amplitudes satisfy the normalization condition $|c_1(0)|^2 + |c_2(0)|^2 = 1$.
The coefficients $c_{1}(0)$ and $c_{2}(0)$ represent the corresponding initial probability amplitudes. Within the single-excitation subspace, the time-dependent state of the composite system and environment evolves according to the ansatz
\begin{equation}\label{statet}
\begin{aligned}
|\psi(t)\rangle = & \left[ c_{1}(t) |e_{c}, g_{B}\rangle + c_{2}(t) |g_{c}, e_{B}\rangle \right] \otimes |0\rangle_{E} \\
& + \sum_{k} c_{k}(t) |g_{c}, g_{B}\rangle \otimes |1_{k}\rangle_{E},
\end{aligned}
\end{equation}
where $|1_{k}\rangle_{E}$ denotes the environmental state with a single photon excitation in mode $k$ and all other modes in the vacuum state. The time-dependent probability amplitudes satisfy the normalization condition $|c_1(t)|^2 + |c_2(t)|^2 + \sum_k |c_k(t)|^2 = 1$.

By substituting the state ansatz Eq.~\eqref{statet} into the time-dependent Schr\"odinger equation $i \frac{d}{dt}|\psi(t)\rangle = H |\psi(t)\rangle$, we obtain the following system of coupled differential equations for the probability amplitudes:
\begin{align}
\dot{c}_1(t) &= -i \omega_0 c_1(t) - i \zeta c_2(t) - i \sum_k \xi_k^{(1)} c_k(t), \label{eq:c1dot} \\
\dot{c}_2(t) &= -i \omega_0 c_2(t) - i \zeta c_1(t) - i \sum_k \xi_k^{(2)} c_k(t), \label{eq:c2dot} \\
\dot{c}_k(t) &= -i \omega_k c_k(t) - i \left[ \xi_k^{(1)*} c_1(t) + \xi_k^{(2)*} c_2(t) \right]. \label{eq:ckdot}
\end{align}

We model the structured environment via a Lorentzian spectral density as $J(\omega) = \frac{\gamma}{2\pi} \frac{\lambda^{2}}{(\omega - \omega_{0})^{2} + \lambda^{2}}$, where $\gamma$ represents the effective system-reservoir coupling strength, while $\lambda$ denotes the spectral bandwidth of the reservoir modes, which sets the reservoir memory timescale as $\tau_E = \lambda^{-1}$. When $\lambda \gg \gamma$, the reservoir loses memory rapidly, leading to the Markovian regime. Conversely, when $\lambda \lesssim \gamma$, memory effects become significant and drive the non-Markovian dynamics. Using the analytical approach developed in Ref.~\cite{b3}, the coupled integro-differential equations for the probability amplitudes can be solved exactly. We express the dynamics in terms of the dimensionless time $\tau = \lambda t$, where $\lambda^{-1}$ is the reservoir correlation time.

We consider the standard charging scenario in which the battery is initially empty, prepared in its ground state and the reservoir is in the vacuum state, while the charger is fully excited, carrying the entire initial excitation of the composite system. To solve the integro-differential equations, we impose the initial conditions as $c_1(0) = 1,  c_2(0) = 0,  c_k(0) = 0  \quad \forall k$,
which correspond to having the charger initially in its excited state, the battery in its ground state, and the reservoir in the vacuum state. The  procedure leads to rather complex expressions for the time-dependent probability amplitudes, which completely determine the excitation dynamics of the quantum charger and the quantum battery. Within the single-excitation subspace, these amplitudes fully determine the state of the battery, allowing us to calculate key performance measures such as the stored energy, ergotropy, and charging power. Furthermore, they enable a detailed investigation into how reservoir memory effects, direct dipole-dipole coupling, and interqubit separation control energy transfer and work extraction.

By performing a partial trace over both the charger and reservoir degrees of freedom in the total state vector in Eq.~\eqref{statet}, the reduced density matrix of the quantum battery is obtained in the bare basis as
\begin{equation}\label{eq:rhoB}
\rho_B(\tau) = |c_2(\tau)|^2 |e_B \rangle \langle e_B| + \left(1 - |c_2(\tau)|^2\right) |g_B \rangle \langle g_B|.
\end{equation}
Using the reduced state of the quantum battery and its free Hamiltonian, and setting the ground-state energy to zero, the internal energy stored in the battery is given by
\begin{equation}\label{eq:EB}
E_B(\tau) = \mathrm{tr}\!\left[\rho_B(\tau) H_B\right] = \omega_0 |c_2(\tau)|^2.
\end{equation}

By substituting the battery reduced density matrix and Hamiltonian into the definition of ergotropy , the extractable work can be expressed in closed form as
\begin{equation}\label{eq:ergotropy_heaviside}
\mathcal{W}(\tau) = \omega_0 \left(2 |c_2(\tau)|^2 - 1\right) \Theta\!\left(|c_2(\tau)|^2 - \frac{1}{2}\right),
\end{equation}
where $\Theta(\cdot)$ denotes the Heaviside step function. This demonstrates that non-vanishing ergotropy strictly requires population inversion, below which the battery state is passive and no work can be extracted via cyclic unitary operations. Furthermore, the instantaneous charging power of the quantum battery is obtained by taking the time derivative of the stored energy in Eq.~\eqref{eq:EB}, which is given by
\begin{equation}\label{eq:inst_power}
\mathcal{P}_B(\tau) =  2\omega_0 \operatorname{Re}\!\big[c_2^*(\tau)\,\dot{c}_2(\tau)\big].
\end{equation}
The overall performance of the charging protocol is evaluated through key performance measures.  Specifically, the maximum stored energy, $\Delta E_B^{\max}$, and the maximum ergotropy, $\mathcal{W}^{\max}$, establish the upper limits for energy storage and work extraction, respectively. Since the battery is initially in its ground state, these quantities are defined as
\begin{equation}\label{eq:max_merits}
\Delta E_B^{\max} = \max_{\tau \ge 0} \Delta E_B(\tau), 
\qquad
\mathcal{W}^{\max} = \max_{\tau \ge 0} \mathcal{W}(\tau),
\end{equation}
where the optimization is carried out over the charging duration $\tau$. Furthermore, the maximum charging power plays a key role in evaluating the dynamical efficiency of the protocol, defined as
\begin{equation}\label{eq:max_power}
\mathcal{P}_B^{\max} = \max_{\tau \ge 0} \mathcal{P}_B(\tau),
\end{equation}
which sets the upper limit on the instantaneous rate of energy transfer to the battery qubit. An ideal charging protocol requires both maximizing $\Delta E_B^{\max}$ and $\mathcal{W}^{\max}$ to ensure substantial energy storage and extractable work, alongside a high maximum charging power $\mathcal{P}_B^{\max}$ to achieve fast charging. These measures provide a comprehensive benchmark for evaluating both the thermodynamic and dynamical performance of the quantum battery. To evaluate how the spatial configuration affects the charging dynamics, we analyze the behavior of these measures in the following.

\begin{figure*}[t]
\centering
\includegraphics[width=0.6\textwidth]{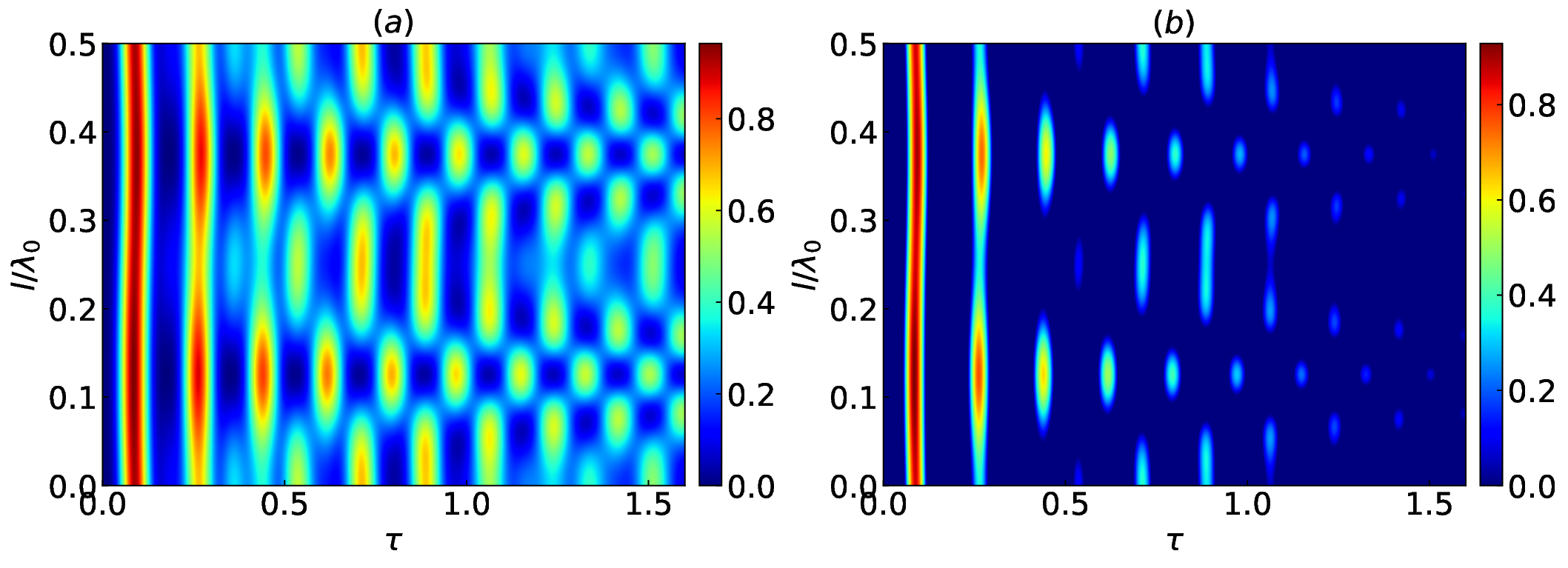}
\hspace{0.05\textwidth}

\caption{\raggedright Time-geometry plot of (a) the internal energy change $\Delta E_B(\tau)$ and (b) the  ergotropy $\mathcal{W}(\tau)$. The battery is initially in its ground state and the charger excited. The parameters are  $\zeta= 0.01\,\omega_0$ and $\lambda/\gamma=0.04$.}\label{fig2}
\end{figure*}

Fig.~\ref{fig2}(a) illustrates the time evolution of the stored energy $\Delta E_B(\tau)$ for different values of the geometric spacing $l/\lambda_0$. Here, the parameters are set to $\zeta = 0.01\,\omega_0$ and $\lambda/\gamma = 0.04$. Since the reservoir spectral width $\lambda$ is much smaller than the coupling rate $\gamma$, the system enters the non-Markovian regime with pronounced memory effects. Consequently, rather than relaxing monotonically, the energy transfer exhibits pronounced coherent backflow whose periodic modulation pattern is directly governed by the spatial separation $l/\lambda_0$. The appearance of distinct vertical bands reveals that the energy transfer efficiency is highly sensitive to the spatial phase. This behavior arises from the dependence of the collective decay rates, $\Gamma_s$ and $\Gamma_a $ on the spatial phase. These rates determine how strongly the symmetric and antisymmetric channels couple to the structured reservoir. As a result, energy transfer is enhanced at some spatial separations and suppressed at others because of constructive and destructive interference. This interplay gives rise to geometry-induced dark configurations, where the charging process is strongly suppressed. The oscillations indicate coherent energy exchange between the charger and the battery. This exchange persists because the reservoir has a long memory time, a characteristic feature of non-Markovian dynamics in the strong-coupling regime. The revival features indicate a partial backflow of energy and information from the reservoir to the battery, which is a characteristic signature of non-Markovian dynamics. Notably, both the maximum stored energy and the optimal charging time depend strongly on the inter-qubit separation. In summary, these results demonstrate that the spatial geometry of the system within a structured environment provides an effective mechanism to control charging efficiency, dynamical stability, and the balance between coherent exchange and dissipation. Specifically, optimal charging occurs at separations that maximize cooperative coupling, whereas destructive spatial interference severely inhibits energy transfer despite strong system–reservoir interactions.



Fig.~\ref{fig2}(b) displays how the battery ergotropy changes with both time and inter-qubit separation, for the same parameters as in Fig.~\ref{fig2}(a). As can be seen, the extractable work depends strongly on both time and inter-qubit separation. Although the battery can store energy across a wide range of parameters, significant ergotropy develops only within specific regions, highlighting the crucial role of the system configuration in work extraction. This indicates that energy storage alone is insufficient for work extraction. The battery must reach a sufficiently non-passive state, and the charger–battery separation also affects the extraction efficiency. Constructive interference enhances the collective coupling and improves energy transfer. This effect leads to higher values of the extractable work, or ergotropy. In contrast, destructive interference suppresses coherent energy exchange and reduces the extractable work across a wide range of parameters.
 The isolated regions of high ergotropy arise from the interplay between coherent dynamics and reservoir memory. In the strong-coupling, non-Markovian regime, partial information backflow can temporarily restore extractable work, producing distinct time windows rather than continuous charging. In summary, the system geometry determines whether stored energy can be extracted as work, setting clear spatial limits on when the ergotropy remains nonzero.
 
\begin{figure*}[t]
\centering
\includegraphics[width=0.3\textwidth]{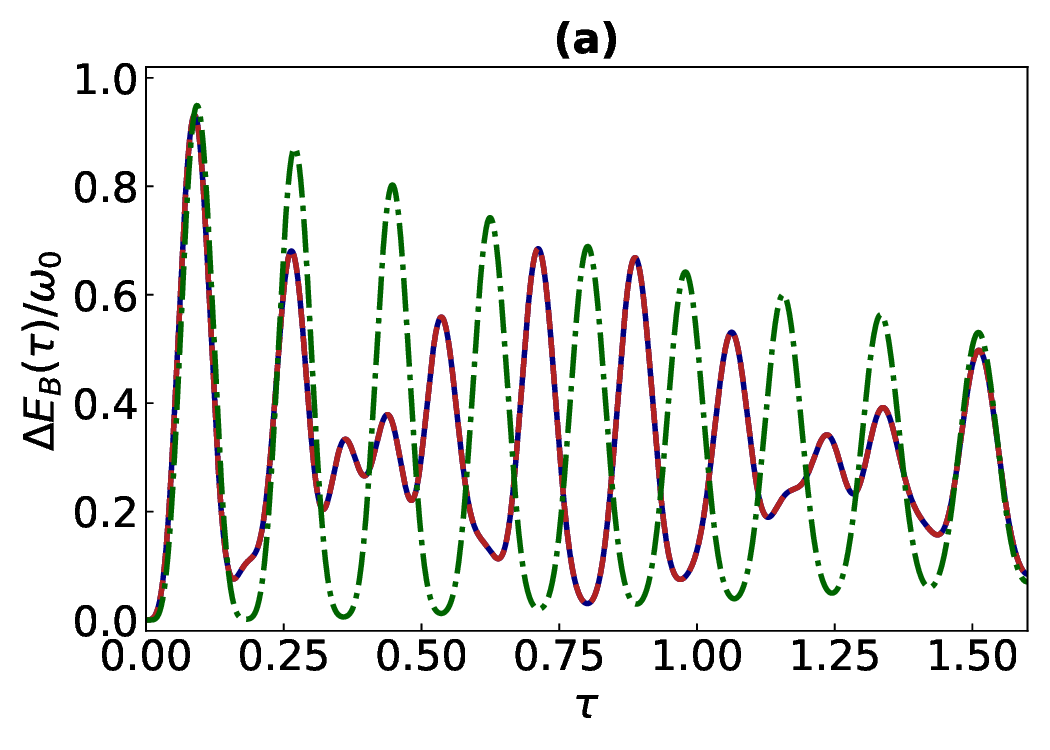}
\hspace{0.02\textwidth}
\includegraphics[width=0.3\textwidth]{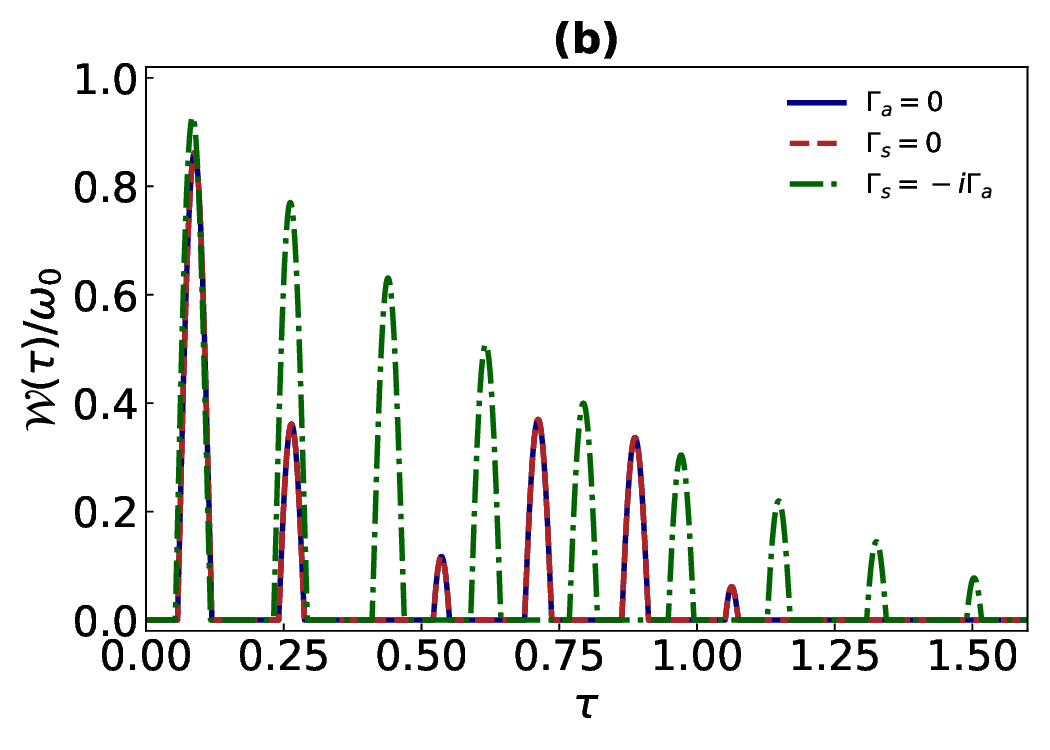}

\caption{\raggedright  (a) The internal energy  $\Delta E_B(\tau)$ and (b) the ergotropy $\mathcal{W}(\tau)$, as function of  $\tau$ for different values of geometry parameter. The battery is initially in its ground state and the charger excited. The parameters are $\zeta=0.01\,\omega_0$, and $\lambda/\gamma=0.04$.}
\label{fig3}
\end{figure*}
Fig.~\ref{fig3}(a) shows how the internal energy changes with time for three representative geometries, defined by $\Gamma_a = 0$, $\Gamma_s = 0$, and $\Gamma_s = -i\Gamma_a$. Here, we also choose the parameters $\zeta=0.01\,\omega_0$ and $\lambda/\gamma=0.04$ to have the strong-coupling, non-Markovian regime. As shown in Fig.~\ref{fig3}(a), the charging dynamics are nearly identical for the $\Gamma_a=0$ and $\Gamma_s=0$ configurations. In both cases, one collective channel is decoupled from the reservoir, while the other remains fully coupled to it.
 Because the initial state has equal symmetric and antisymmetric components, suppressing either channel leaves one active collective mode, with a similar effective coupling strength in both cases.
 As a result, exchanging the two channels leaves the internal-energy dynamics unchanged, giving identical charging behavior. By contrast, in the mixed configuration, $\Gamma_s=-i\Gamma_a$, both collective channels couple to the reservoir with equal strengths.
 In this regime, both symmetric and antisymmetric components contribute to the dynamics, enhancing energy exchange between the charger and the battery. This combined contribution increases the oscillation amplitude and improves the charging performance. The long-lived oscillations in all configurations result from the narrow reservoir spectral width.
 This condition enhances reservoir memory and allows energy to be exchanged repeatedly between the quantum battery and the structured reservoir. The main result of Fig.~\ref{fig3}(a) is that the battery stores the most energy when both collective channels contribute coherently. The single-channel configurations show lower stored energy, even though the active channel remains fully coupled to the reservoir.

Figure~\ref{fig3}(b) shows how the ergotropy evolves over time for the same different geometric configurations examined in Figure~\ref{fig3}(a). The purely symmetric and purely antisymmetric configurations show nearly identical ergotropy dynamics. In each case, one collective channel contributes to the evolution, while the other is suppressed. Since the initial state contains equal contributions from the two collective components, suppressing either channel leads to the same effective dynamics and therefore the same work-extraction behavior. In comparison with the purely symmetric and purely antisymmetric configurations, the mixed configuration produces higher ergotropy and more pronounced revivals. When both collective channels contribute, their coherent interaction increases energy transfer to the battery and raises its population inversion, resulting in greater extractable work. The recurring local maxima in ergotropy reflect the non-Markovian nature of the dynamics, which allow energy to be exchanged repeatedly between the quantum battery and its environment. These results show that work extraction is enhanced when the symmetric and antisymmetric channels contribute jointly, demonstrating the role of geometry in quantum battery performance.

\begin{figure*}[t]
\centering
\includegraphics[width=0.75 \textwidth]{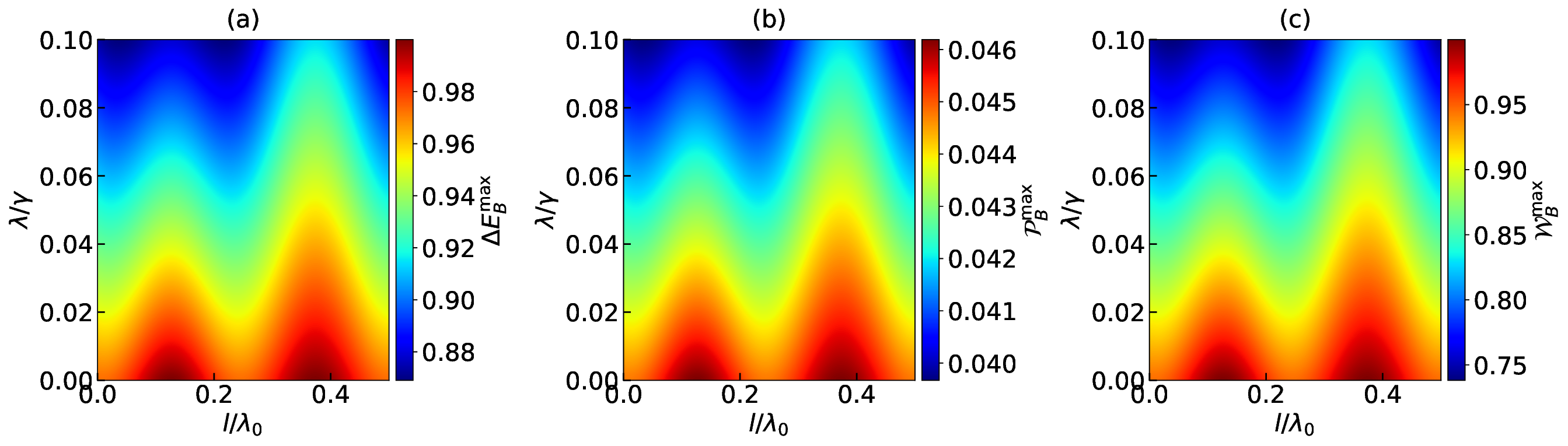}
\hspace{\textwidth}
\caption{\raggedright  (a) Maximum internal energy $\Delta E_B^{\max}$, (b) maximum charging power $\mathcal{P}_B^{\max}$, and (c) maximum ergotropy $\mathcal{W}_B^{\max}$ as functions of the geometric inter-qubit separation parameter $l /\lambda_0$ and  $\lambda/\gamma$. 
}
\label{fig4}
\end{figure*}
Fig.~\ref{fig4} shows how the quantum battery performance depends on the geometric parameter and the reservoir spectral width. Fig.~\ref{fig4}(a) shows the maximum internal energy, $\Delta E_B^{\max}$, while Fig.~\ref{fig4}(b) and Fig.\ref{fig4}(c) show the maximum charging power $\mathcal{P}_B^{\max}$ and maximum ergotropy $\mathcal{W}_B^{\max}$, respectively, as functions of the geometric inter-qubit separation parameter and the normalized spectral width. In all three plot, battery performance is enhanced when the normalized spectral width $\lambda/\gamma$ is small, where reservoir memory effects are strong. As the normalized spectral width increases, the maximum stored energy and ergotropy decrease. This trend reflects the gradual weakening of coherent energy exchange as the system moves toward the Markovian regime. This reduction reflects weaker information backflow and a gradual shift toward more irreversible energy dissipation.

Along the geometric axis, all three quantities exhibit a clear periodic modulation. Along the geometric axis, all three quantities vary periodically with the separation. Their maximum value occur at specific separations where collective interference enhances the contribution of the symmetric and antisymmetric channels to the dynamics. In contrast, performance is reduced near configurations involving partial destructive interference. The optimal geometric configurations are similar for internal energy, charging power, and ergotropy, showing that geometry can control both energy storage and work extraction. In summary, these results show that the best battery performance requires both strong collective interference, achieved by tuning the spatial phase, and long reservoir memory, corresponding to small $\lambda/\gamma$. This shows that geometric design and non-Markovian dynamics can jointly improve energy storage and work extraction.

\begin{figure*}[t]
\centering
\includegraphics[width=0.7 \textwidth]{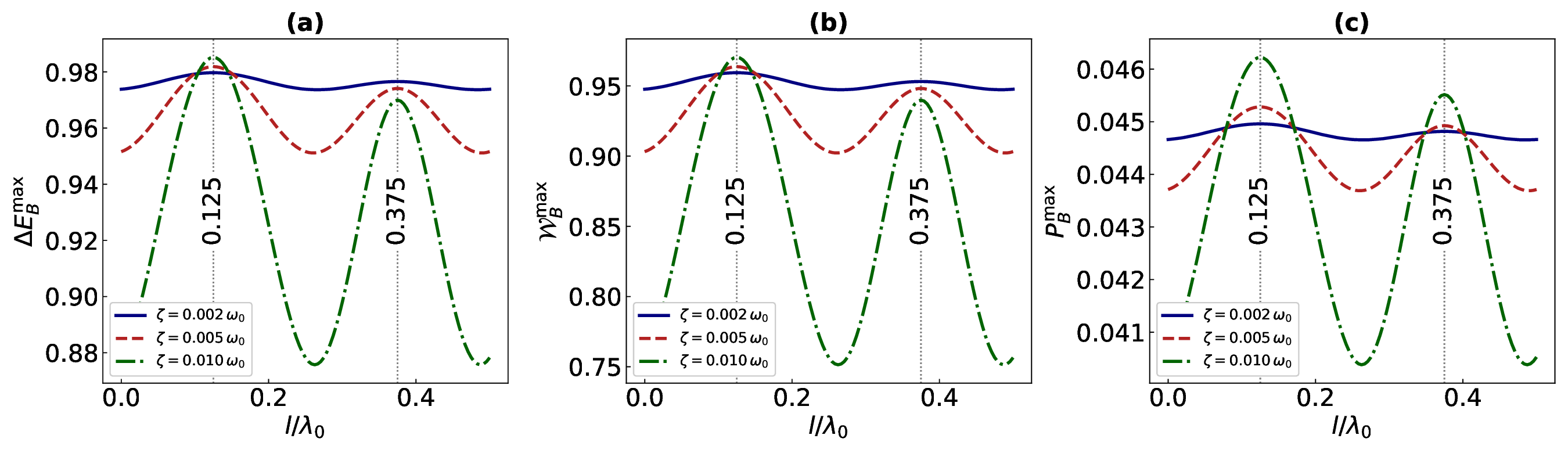}
\hspace{\textwidth}
\caption{(Color online) (a) Maximum internal energy $\Delta E_B^{\max}$, (b) maximum ergotropy $\mathcal{W}_B^{\max}$, and (c) maximum charging power $\mathcal{P}_B^{\max}$ as a function of the inter-qubit separation $l/\lambda_0$  for $\zeta = 0.002\,\omega_0$ (solid navy), $\zeta = 0.005\,\omega_0$ (dashed red), and $\zeta = 0.010\,\omega_0$ (dash-dotted green) with $\lambda/\gamma = 0.04$. Vertical dotted lines denote the optimal distance configurations at $l/\lambda_0 = 0.125$ and $0.375$.}
\label{fig5}
\end{figure*}


Fig.~\ref{fig5} shows the maximum internal energy $\Delta E_B^{\max}$, ergotropy $\mathcal{W}_B^{\max}$ and charging power $P_B^{\max}$ as functions of the normalized inter-qubit separation $l/\lambda_0$ and the direct coherent coupling $\zeta$ in subplots (a), (b), and (c), respectively.  

 Fig.~\ref{fig5}(a) displays the maximum stored energy in the battery, $\Delta E_B^{\max}$, as a function of the normalized separation $l/\lambda_0$ for three representative values of the direct coherent coupling $\zeta$. The stored energy exhibits a periodic modulation governed by the spatial phase $\theta$, showing pronounced maximum at $l/\lambda_0 = 0.125$ and $l/\lambda_0 = 0.375$, corresponding to $\theta = \pi/4$ and $\theta = 3\pi/4$, respectively, and distinct minimum at $l/\lambda_0 = 0.25$ corresponding to $\theta = \pi/2$, as well as at the boundary values $l/\lambda_0 = 0$ and $0.5$. At $l/\lambda_0 = 0.125$, the symmetric and antisymmetric collective coupling strengths balance and interfere constructively with the direct coupling channel $\zeta$, leading to optimal coherent population transfer to the battery mode. Conversely, near $l/\lambda_0 = 0.25$, destructive interference suppresses complete energy transfer and reduces the charging capacity. Furthermore, the sensitivity of the system to the spatial phase becomes significantly more pronounced as the direct coherent coupling $J$ increases. For a weak coupling of $J = 0.002\,\omega_0$, the maximum stored energy $\Delta E_B^{\max}$ remains relatively stable across different positions with variations on the order of $0.005\,\omega_0$. In contrast, when the coupling strength is increased to $\zeta = 0.010\,\omega_0$, the modulation amplitude expands drastically.

Fig.~\ref{fig5}(b) shows the maximum ergotropy, $\mathcal{W}_B^{\max}$, as a function of $l/\lambda_0$.  The maximum ergotropy reaches its maximum value at the optimal separation $l/\lambda_0 = 0.125$ for $\zeta = 0.010\,\omega_0$, but undergoes a substantial drop at $l/\lambda_0 = 0.25$. This sharp reduction indicates that destructive interference caused by the inter-qubit separation degrades not only the raw energy stored in the battery, but also the extractable work. 

Finally, Fig.~\ref{fig5}(c) illustrates the maximum charging power, $P_B^{\max}$, versus $l/\lambda_0$. Consistent with the energy and ergotropy dynamics, the highest value of the maximum charging power is achieved at $l/\lambda_0 = 0.125$ for the largest direct coupling $\zeta = 0.010\,\omega_0$. The spatial tuning of $l/\lambda_0$ enables a significant speedup in charging rates around constructive interference points ($0.125$ and $0.375$), demonstrating that precise geometric placement of the qubits within the structured waveguide serves as an effective control parameter to simultaneously optimize charging capacity, work extraction, and charging rate.

\section{Conclusion}\label{sec:conclusion}
In this work, we explored how the spatial configuration of a two-qubit quantum battery embedded in a structured waveguide can be used as an active control mechanism for charging performance and work extraction. By examining the interplay between direct inter-qubit tunneling, collective waveguide-mediated decay, and structured reservoir memory, we demonstrated that the spatial phase acquired across the qubit separation fundamentally dictates the energy-transfer pathways. Our findings reveal that charging is not governed merely by the coupling strength, but by wave interference between symmetric and antisymmetric collective decay channels. At specific spatial separations, constructive interference opens balanced conduits that dramatically enhance energy transfer, peak charging rates, and extractable work. Conversely, destructive interference suppresses these channels, leading to quasi-dark states where excitations remain trapped or dissipate without effectively charging the battery. Crucially, the thermodynamic quality of the stored energy, quantified by ergotropy, displays a much stricter sensitivity to spatial positioning than the total stored energy. Because work extraction demands population inversion in the battery, even modest destructive interference can sharply eliminate extractable work while raw energy is still present. Furthermore, reservoir memory effects and coherent inter-qubit coupling act synergistically with spatial phase control. Strong non-Markovian memory sustains coherent energy exchange and drives robust ergotropy revivals, but this environmental backflow can only be fully harnessed when the spatial phase is matched constructively. Increasing the direct coupling between the charger and the battery accentuates this contrast, turning the inter-qubit separation into a sensitive geometric dial. Overall, our study demonstrates that spatial phase engineering offers an accessible and powerful design principle for open quantum batteries. Rather than treating waveguide dissipation purely as a source of decoherence, structured interference allows the environment to be tailored into a resource that simultaneously boosts energy capacity, thermodynamic work output, and charging speed in waveguide-based quantum technologies.


\end{document}